\documentclass[prl,showpacs,amsmath,amssymb,superscriptaddress, twocolumn,letterpaper]{revtex4}

\usepackage{amssymb}
\usepackage{graphicx}
\usepackage{color}
\usepackage{dcolumn}
\usepackage{bm}

\newcommand \T{\rule{0pt}{2.6ex}}

\begin{document}

\title{High-pressure study of non-Fermi liquid and spin-glass-like behavior in CeRhSn}

\author{D. A. Zocco}
\altaffiliation[Present address: ]{Institut f\"{u}r Festk\"{o}rperphysik, Karlsruhe Institute of Technology, D-76021 Karlsruhe, Germany.}
\affiliation{Department of Physics, University of California, San Diego, La Jolla, California 92093, USA}

\author{A.~\'{S}lebarski}
\affiliation{Institute of Physics, University of Silesia, 40-007 Katowice, Poland}

\author{M. B. Maple}
\affiliation{Department of Physics, University of California, San Diego, La Jolla, California 92093, USA}

\date{\today}

\begin{abstract}
We present measurements of the temperature dependence of electrical resistivity $\rho(T)$ of CeRhSn up to $\sim$ 27 kbar. At low temperatures, $\rho(T)$ varies linearly with $T$ for all pressures, indicating non-Fermi liquid behavior. Below $T_f \sim$ 6 K, $\rho(T)$ deviates from a linear dependence. We found that the low-$T$ feature centered at $T=T_f$ shows a pressure dependence $\partial T_f/\partial P \approx 30$ mK/kbar which is typical of canonical spin glasses. This interplay between spin-glass-like and non-Fermi liquid behavior was observed in both CeRhSn and a Ce$_{0.9}$La$_{0.1}$RhSn alloy.
\end{abstract}

\pacs{62.50.-p, 71.27.+a, 71.30.+h, 72.15.Qm}

\maketitle

\newpage

\section{Introduction}

The low-temperature anomalous behavior observed in many heavy-fermion (HF) systems has frequently been attributed to the proximity of a quantum critical point (QCP). A QCP refers to the value of a control parameter, such as pressure \cite{Mathur98}, chemical composition \cite{Loh94}, or magnetic field \cite{Heuser98}, where a second order phase transition is suppressed to 0 K. Starting from a long-range ordered state, the suppression of magnetic order by varying the control parameter could signal a 0 K quantum phase transition. For such a magnetic-nonmagnetic transition, strong deviations from Fermi liquid behavior are expected to occur \cite{Hertz76,Continentino93,Millis94}. Recently, we have observed non-Fermi liquid (NFL) behavior in polycrystalline CeRhSn \cite{Sle02,Sle02a}. The electrical resistivity $\rho$, magnetic susceptibility $\chi$, and specific heat $C/T$, showed power-law temperature dependences: $\rho (T) \sim T$ and $\chi (T) \sim C(T)/T \sim T^{-0.5}$. The NFL behavior was speculated to originate from Griffiths singularities as a consequence of the interplay between an intrasite Kondo effect and intersite Ruderman-Kittel-Kasuya-Yosida (RKKY) interaction in the presence of disorder and magnetic anisotropy \cite{Sle02}. Very similar power law dependences of $\rho (T)$, $\chi (T)$ and $C(T)/T$ were later reported for a single crystal, but with a strongly anisotropic electrical resistivity \cite{Kim03}. The results obtained for polycrystalline samples are consistent with a mixture of the $T$-dependences observed in the single crystals.

Cerium-based Kondo-lattice systems exhibit a variety of exotic ground states, including heavy fermion and non-Fermi liquid behavior in the metallic state, or Kondo insulating behavior \cite{tsunetsugu1997}. The heavy fermion Fermi liquid state in the intermetallic $f$-electron compounds can be modeled via the periodic Anderson model (PAM) for both paramagnetic and magnetic cases \cite{And61}. The stability of paramagnetic (PM), ferromagnetic (FM) and antiferromagnetic (AFM) states in the Kondo-lattice limit was recently discussed by the theory of Doradzi\'{n}ski and Spa\l ek (DS) \cite{Spa97}. This theory describe in great detail the ground state properties of ternary Ce-based intermetallics of the form Ce$MX$ ($M$ = Rh, Pd or Ni, and $X$ = Sb, Sn or Al) in terms of the total number of electrons per site $n_e$ and the strength of the hybridization potential $V_{fc}$ that admixes the Ce 4$f$-electron states and the conduction electron states \cite{Sle06}. In the $n_e-V_{fc}$ diagram, CeRhSn has been identified as a possible AFM metal or an AFM Kondo insulator (AKI) based on estimates of $n_e \sim$ 2 and $V_{fc} \sim$ 150 meV \cite{note}. In this case, the DS model predicts the transformation from an AKI to a paramagnetic Kondo insulator (PKI) at higher values of $V_{fc}$ expected at high pressures. While CeRhSn displays metallic behavior down to the lowest measured temperatures, the proximity of CeRhSn to a Kondo insulating state has been also inferred from Rh-doping studies of the CeNiSn Kondo insulating material \cite{sle2008, sle2009}. Although no signs of AFM order have been found in poly- or single-crystalline samples, $^{119}$Sn nuclear magnetic resonance (NMR) experiments suggest that CeRhSn is located in the vicinity of an AFM instability \cite{tou2004}. CeRhSn could also be interpreted as a semi-metallic Kondo-lattice system with a pseudogap in the density of states (DOS) at the Fermi level and spin-glass-type ordering (rather than AFM ordering).

Motivated by the predictions of the DS model for the Ce-based intermetallic ternary compounds and the above mentioned evidence of NFL behavior, we studied the effects of externally applied hydrostatic pressure to polycrystalline samples of CeRhSn and its alloy Ce$_{0.9}$La$_{0.1}$RhSn via measurements of their electrical resistivity. The experiments have shown no indication of a AKI-PKI transition. Pressure-dependent interplay of NFL behavior and spin-glass-like behavior was observed in both CeRhSn and Ce$_{0.9}$La$_{0.1}$RhSn alloys.

\begin{figure*}
\begin{center}
{\includegraphics[width=6.5in]{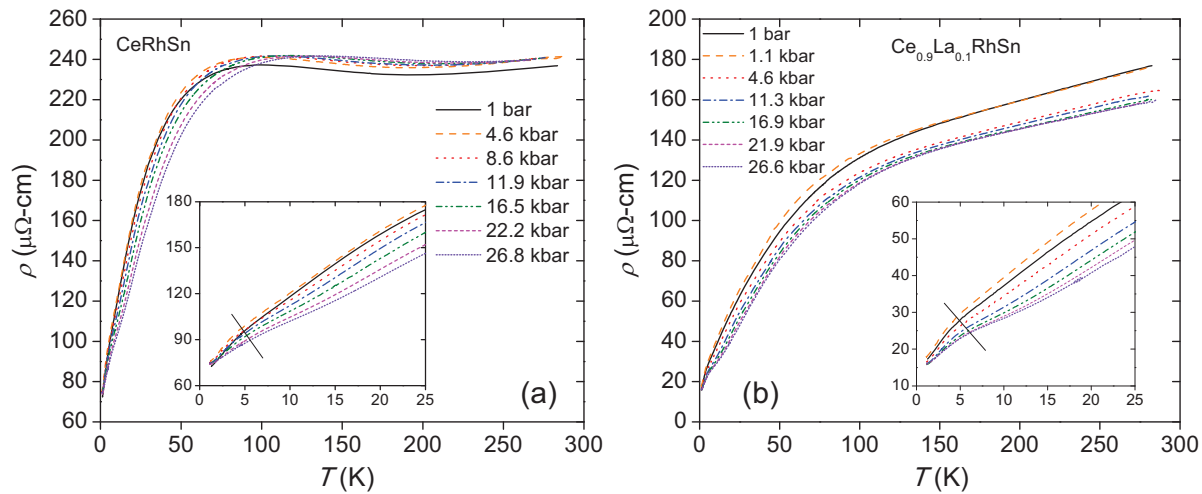}}
\end{center} \caption{Electrical resistivity $\rho(T)$ versus temperature $T$ at different external pressures for (a) CeRhSn and (b) Ce$_{0.9}$La$_{0.1}$RhSn. The \textit{insets} show the broad feature at $T_f$ below 10 K.} \label{fig:Fig1}
\end{figure*}

\section{Experimental details}

Polycrystalline samples of CeRhSn and Ce$_{0.9}$La$_{0.1}$RhSn were prepared by arc melting the constituent elements (Ce 99.99\%, La 99.9\%, Rh 99.9\%, Sn 99.999\% in purity) on a water cooled copper hearth in an argon atmosphere with a Zr getter. The melting process was repeated several times to promote homogeneity and the resultant ingots were annealed at 800$^\circ$C for 2 weeks and then quenched in water. The samples were examined by x-ray diffraction and found to consist of a single phase, crystallizing in a hexagonal unit cell of the Fe$_{2}$P-type structure (space group P$\bar{6}2m$).

Electrical resistivity measurements under pressure were performed in a beryllium-copper, piston-cylinder clamped cell. A 1:1 mixture of n-pentane and isoamyl alcohol in a Teflon capsule served as the pressure transmitting medium to ensure hydrostatic conditions during pressurization at room temperature. The pressure in the sample chamber was inferred from the inductively determined, pressure-dependent superconducting critical temperature of a tin manometer \cite{smith69}. From the widths of these transitions, we estimated pressure gradients as large as 3\% of each measured pressure. Electrical contacts were made with 50 $\mu$m-gold wire attached to the samples with silver epoxy and cured at 200$^\circ$C for five minutes. In all cases, the electrical resistance was measured using a four-lead technique and a Linear Research Inc.\ LR-700 ac resistance bridge, with excitations smaller than 1 mA.

\section{Results}

\begin{figure}
{\includegraphics[width=3.2in]{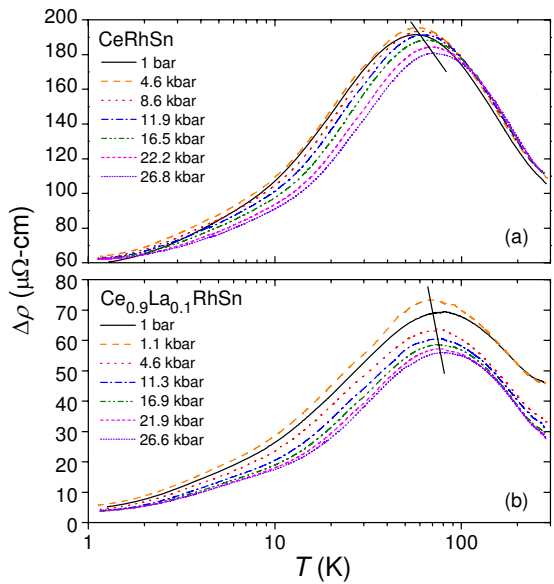}} \caption{Magnetic contribution to the electrical resistivity $\Delta \rho (T) = \rho
(CeRhSn,T) - \rho (LaRhSn,T)$ versus $ln(T)$ measured at different external pressures $P$ for (a) CeRhSn and for (b) Ce$_{0.9}$La$_{0.1}$RhSn.} \label{fig:Fig2}
\end{figure}

Displayed in figure \ref{fig:Fig1} are the electrical resistivity $\rho(T)$ data for CeRhSn and Ce$_{0.9}$La$_{0.1}$RhSn, at different values of pressure $P$ between 1 bar and 27 kbar \cite{note}. In this pressure range, local maxima are found between 90 and 150 K in CeRhSn, which appear as broad shoulders around 100 K in the La-substituted sample. In order to analyze these features, we plotted $\Delta \rho(T) = \rho(CeRhSn, T) - \rho(LaRhSn, T)$ (\textit{upper panel}) and $\Delta \rho(T) = \rho(Ce_{0.9}La_{0.1}RhSn, T) - \rho(LaRhSn, T)$ (\textit{lower panel}) in figure \ref{fig:Fig2}. Although the resistivity of Ce$_{0.9}$La$_{0.1}$RhSn does not exhibit coherence maxima, the effect of coherence is clearly evident as a broad shoulder in the resistivity of the doped sample at similar temperatures. These $P$-dependent maxima at $T_{coh} \sim$ 60 - 80 K and the $\Delta \rho \sim -ln(T)$ behavior at higher temperatures shown in figure \ref{fig:Fig2} provide evidence that the resistivity maxima result from a competition between quantum coherence ($i.e.$ itineracy of 4$f$-electrons due to the hybridization of the localized $f$-electron states with the conduction electron states) and the thermal disorder acting as a decoherence factor. From figure \ref{fig:Fig2}, we can also estimate the pressure dependence of $T_{coh}$, obtaining $\partial T_{coh}/\partial P \approx 0.65$ K/kbar for CeRhSn. This result suggests that the hybridization strength between the localized $f$- and conduction-electron states increases continuously with pressure. The pressure dependence of $T_{coh}$ in $\Delta \rho (T)$ for Ce$_{0.9}$La$_{0.1}$RhSn is also linear in applied pressure, and $\partial T_{coh}/\partial P \approx 0.4$ K/kbar. The temperature $T_{coh}$ characterizes experimentally the ``effective'' Kondo (hybridization) temperature $T_K$. The Anderson model \cite{And61} predicts that $k_B T_{coh} \sim \pi\langle V_{fc}^{2}\rangle N(E_F)$, where $\langle V_{fc} \rangle$ is the matrix element that admixes the Ce 4$f$-electron states and the conduction electron states, and $N(E_F)$ is the density of electronic states at the Fermi energy. Assuming $N(E_F)\approx (1/2) $ state/(eV atom) and $\langle V_{fc} \rangle \sim 0.1$ eV \cite{Sle02,Gamza09}, we obtained $T_{coh} \sim$ 150 K, a correct order of magnitude.

\begin{figure}
{\includegraphics[width=3.2in]{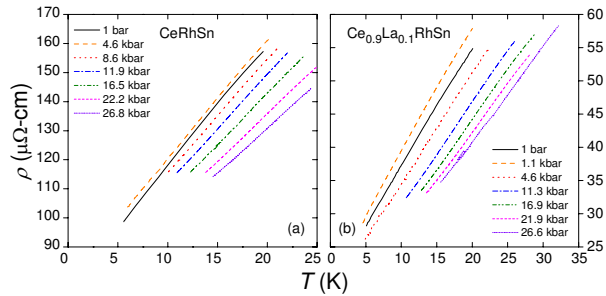}} \caption{Electrical resistivity $\rho$ as a function of $T$ between $T_f$ and 30 K. The resistivity $\rho$ is well fitted by the expression $\rho (T) = A+BT^n$; $n \approx 1$ for (a) CeRhSn and also for (b) Ce$_{0.9}$La$_{0.1}$RhSn.} \label{fig:Fig3}
\end{figure}

At temperatures below $T_{coh}$, the electrical resistivity displays a linear dependence with temperature down to $T_f$ $\sim$ 6 K, where a broad feature develops (see the \textit{insets} of figure \ref{fig:Fig1}). In figures \ref{fig:Fig3} (a) and (b), $\rho (T,P)$ is plotted below 30 K, in order to emphasize its linear behavior. Table I groups the results of the fits of these curves with the equation $\rho (T) = \rho_0 +AT^n$ in the temperature range $T_f <T\lesssim$ 30 K. For both compounds, we found that $n \approx$ 1 in this temperature range at all pressures. The deviation of $\rho (T)$ from linear behavior below $T_f$ could be possibly originating from the inhomogeneous magnetic ordering of spin-glass-type, promoted by atomic disorder \cite{sle2009b}. The spin glass-like behavior was previously observed for the Ce$_{1-x}$La$_{x}$RhSn compounds below $T_f$ \cite{Sle02a,sle2004}. In that work, we concluded that the spin-glass-like mechanism could be responsible for the breakdown of the divergent behavior in the specific heat $C/T $ of CeRhSn, which saturates below $\sim$ 1 K \cite{Sle02,Kim03}. Our high-pressure experiments show that the characteristic temperature of these features increase with pressure at the rate $\partial T_f/\partial P \approx 30$ mK/kbar, a typical value for spin glasses \cite{Hard80,Sch79}.

\section{Discussion}

It has recently been shown that a magnetic phase diagram \cite{Spa97} on the $V_{fc}-n_e$ plane describes reasonably well the ground-state properties of a series of ternary Ce-compounds \cite{Sle06}, where $n_e = n_c + n_f$ is the total number of electrons per site, with $n_c$ and $n_f$ being the total number of conduction and localized electrons, respectively. In this diagram, CeRhSn has been identified as a possible antiferromagnetic metal or even an antiferromagnetic Kondo insulator (AKI) based on estimates of $n_e\approx 2$ and $V_{fc}\approx 150$ meV \cite{note2}. For the AKI alternative, CeRhSn could be interpreted as a semimetallic weakly magnetic Kondo-lattice system with a pseudogap in the density of states (DOS) at the Fermi level. A pseudogap in the DOSs of CeRhSn is expected from recent calculations \cite{Sle02,Gamza09}. In the mean-field solution \cite{Spa97}, the Kondo-like compensation of the Ce magnetic moments is not complete for the AKI phase, and the small value of the magnetic moment results from the localization of the $f$-states expressed by $n_f \rightarrow 1$. The periodic Anderson model predicts the transformation from an AKI to a paramagnetic Kondo insulator (PKI) at $V_{fc}\lesssim$ 350 meV \cite{Spa97}. As we mentioned in the previous section, in reality CeRhSn is not an intrinsic antiferromagnet nor an insulator, but exhibits an inhomogeneous magnetic state of a spin-glass-type, and calculations have shown that a pseudogap exists in the density of states at the Fermi level \cite{Sle02a}.

In the periodic Anderson model, the magnetic/nonmagnetic behavior of a dense Kondo system is controlled by the strength of the $f-c$ hybridization matrix element $V_{fc}$ between the localized $f$-electron and conduction-electron states \cite{And61,Spa97}. Application of pressure is also known \cite{Sch79,Maple74} to increase the value of $\left|JN(\epsilon_F)\right|$ in Ce compounds ($J$ is the exchange interaction parameter). With increasing pressure, therefore, quantum critical behavior should also be observed in magnetic Ce-compounds when the ordering temperature is suppressed to 0 K \cite{Don77}. As noted above, the transition from a weakly magnetic phase of CeRhSn to a PKI state has been expected to occur under high pressure, although this does not seem to be the case, at least in the pressure range covered by our experiments.
On the other hand, our low-temperature electrical resistivity versus pressure data confirm the critical behavior of CeRhSn, where we found that $\rho(T) \sim T$ (see Table I). The quantum coherence temperature $T_{coh}$ is, however, strongly pressure dependent ($\partial T_{coh}/\partial P \approx 0.65$ K/kbar), which is characteristic of many Ce-based compounds. A large $P$-dependence of $T_{coh}$ seems to be a general feature of heavy fermion systems \cite{Mederle02}, since the Kondo temperature is very sensitive to pressure. Lanthanum substitution in CeRhSn increases the density of states $N(\epsilon_F)$ \cite{Sle02a}. As a consequence, a Ce$_{0.9}$La$_{0.1}$RhSn alloy shows more metallic character in the resistivity than the parent compound. The coherence maximum is nevertheless still observed in the corresponding $\Delta \rho (T)$ curves, but the atomic disorder leads to a smaller value of $\partial T_{coh}/\partial P \approx$ 0.4 K/kbar in this alloy. For both compounds, the coefficient $A$ listed in Table I was found to decrease monotonically in the high pressure experiments ($P >$ 1 bar). For Kondo compounds which display Fermi liquid behavior at $T \ll T_K$, the low-temperature electrical resistivity can be described as $\rho(T) = \rho_0 + A'T^2$, with $A'$ approximately proportional to $T^{-2}_{K}$. For non-Fermi liquid systems like CeRhSn, however, the relation between the coefficient $A$ and $T_K$ is not necessarily valid anymore, and one could only state that a decrease of the coefficient $A$ can be related with an increase of a characteristic temperature of the non-Fermi liquid state, usually related to spin or spatial fluctuation energy scale \cite{oomi2004}.

The residual resistivity $\rho_0$ of CeRhSn and Ce$_{0.9}$La$_{0.1}$RhSn decreases continuously with increasing $P$; for CeRhSn, $\rho_0$ is reduced by $\Delta\rho_0\approx$ 10 $\mu \Omega cm$ at $P$ = 26.8 kbar relative to the value of $\rho_0$ at low pressure. If the residual resistivity arises from coherent spin fluctuations, then the application of a sufficiently large magnetic field would suppress the spin fluctuations and depress the resistivity significantly. The residual resistivity of CeRhSn measured in high magnetic fields up to 18 T increases, however, with increasing magnetic field \cite{Kim03}. This observation suggests that an additional scattering mechanism originating from magnetic moments operates below 10 K \cite{Kim03}. In figure \ref{fig:Fig1} (a) and (b), the electrical resistivity exhibits a broad and weak feature at $T_f$, which displays a positive shift of about $\partial T_f/\partial P \approx$ 30 mK/kbar with increasing pressure, that is characteristic of many well known classical spin glasses, \textit{e.g.} Au:Fe, Ag:Mn, La:Ce \cite{Hard80,Sch79}. Our recent magnetic studies \cite{Sle02a} have suggested the presence of superparamagnetic clusters in CeRhSn, originating from disorder. The cluster-model explains qualitatively the nature of the low-$T$ temperature behavior $\chi (T)\sim C(T)/T \sim T^{-n}$ in CeRhSn with $n \approx 0.5$. We suggest that a magnetic ground state of CeRhSn could be responsible for the existence of a Griffiths-McCoy phase \cite{Griffiths69}. This model predicts power-law behavior of $C(T)/T$ and $\chi(T)$ with similar power-law exponents.

\begin{table}
\caption{\label{tab:Table1}
Parameters obtained from the fits to the electrical resistivity of CeRhSn and Ce$_{0.9}$La$_{0.1}$RhSn by the relation $\rho=\rho_0 + AT^n$ below 30 K (from Fig.~\ref{fig:Fig3}).}
\begin{ruledtabular}
\begin{tabular}{cccc}

       &$P$ (kbar)& $\rho=\rho_0 + AT^n$ \\
       &          & $\rho_0$ ($\mu\Omega$cm)~~~$A$ ($\mu\Omega$cm/K)~~~$n$\\
\hline

CeRhSn \T & 0.001  & ~~~~~~~75.4~~~~~~~~~~~~4.26~~~~~~~~0.89\\
       & 4.6  & ~~~~~~~79.1~~~~~~~~~~~~4.13~~~~~~~~0.96\\
       & 8.6   & ~~~~~~~76.3~~~~~~~~~~~~3.92~~~~~~~0.94\\
       & 11.9  & ~~~~~~~74.2~~~~~~~~~~~~3.76~~~~~~~~0.99\\
       & 16.5  & ~~~~~~~72.2~~~~~~~~~~~~3.53~~~~~~~~0.99\\
       & 22.2  & ~~~~~~~69.8~~~~~~~~~~~~3.30~~~~~~~~1.01\\
       & 26.8  & ~~~~~~~68.5~~~~~~~~~~~~3.12~~~~~~~~1.12\\
\hline

Ce$_{0.9}$La$_{0.1}$RhSn \T & 0.001     & ~~~~~~~19.3~~~~~~~~~~~~1.79~~~~~~~~0.91\\
                            & 1.1   & ~~~~~~~20.4~~~~~~~~~~~~1.90~~~~~~~~0.88\\
                            & 4.6   & ~~~~~~~17.9~~~~~~~~~~~~1.66~~~~~~~~1.00\\
                            & 11.3  & ~~~~~~~15.4~~~~~~~~~~~~1.57~~~~~~~~1.01\\
                            & 16.9  & ~~~~~~~14.0~~~~~~~~~~~~1.51~~~~~~~~1.00\\
                            & 21.9  & ~~~~~~~13.2~~~~~~~~~~~~1.46~~~~~~~~1.09\\
                            & 26.6  & ~~~~~~~12.4~~~~~~~~~~~~1.43~~~~~~~~1.15\\

\end{tabular}
\end{ruledtabular}
\end{table}

In summary, we performed high-pressure electrical resistivity measurements on CeRhSn polycrystals and its lanthanum substituted variant Ce$_{0.9}$La$_{0.1}$RhSn. We found that the features associated with quantum coherence move to higher temperatures due to the effect of pressure. No evidence of the predicted transition from an antiferromagnetic Kondo insulator (AKI) to a paramagnetic Kondo insulator (PKI) was found in our experiments up to 26.8 kbar, although higher pressures might be necessary in order to observe it. We also found that in both compounds, $\rho (T) \sim T$ in the range $T_f < T \lesssim$ 30 K, and this non-Fermi liquid behavior with power-law exponent $n \approx$ 1 breaks down below $T_f \approx$ 6 K. The increase in the electrical resistivity at lower temperatures was previously explained as possibly originating from inhomogeneous magnetic ordering of the spin-glass-type, promoted by atomic disorder. We observed that this low-temperature feature increases with pressure at 30 mK/kbar, which is equivalent to the rates found in typical spin-glasses.

\section{Acknowledgments}

High-pressure research at University of California, San Diego, was supported by the National Nuclear Security Administration under the Stewardship Science Academic Alliance program through the U.S. Department of Energy grant number DE-52-09NA29459. D.Z. gratefully acknowledge discussions with K. Grube and J. Hamlin. A.\'S. was supported by the Ministry of Science and Higher Education within the research project No N N202 032137.


\begin{thebibliography}{99}

\bibitem{Mathur98}
Mathur N D, Grosche F M, Julian S R, Walker I R, Freye D M, Haselwimmer R K W and Lonzarich G G 1998 {\it Nature} {\bf 394} 39

\bibitem{Loh94}
L\"ohneysen H v, Pietrus T, Portisch G, Schlager H G, Schr\"{o}der H, Sieck M and Trappmann T G 1994 {\it Phys. Rev. Lett.} {\bf 72} 3262

\bibitem{Heuser98}
Heuser K, Scheidt E -W, Schreiner T and Stewart G R 1998 {\it Phys. Rev. B} {\bf 57} R4198

\bibitem{Hertz76}
Hertz J A 1976 {\it Phys. Rev. B} {\bf 14} 1165

\bibitem{Continentino93}
Continentino M A 1993 {\it Phys. Rev. B} {\bf 47} 11587

\bibitem{Millis94}
Millis A J 1993 {\it Phys. Rev. B} {\bf 48} 7183

\bibitem{Sle02}
\'Slebarski A, Maple M B, Freeman E J, Sirvent C, Rad\l owska M, Jezierski A, Granado E, Huang Q and Lynn J W 2002 {\it Phil. Mag. B} {\bf 82} 943

\bibitem{Sle02a}
\'Slebarski A, Rad\l owska M, Zawada T, Maple M B, Jezierski A and Zygmunt A 2002 {\it Phys. Rev. B} {\bf 66} 104434

\bibitem{sle2004}
\'Slebarski A, Grube K, Lortz R, Meingast C and L\"ohneysen H v 2004 {\it J. Magn. Magn. Mater.} {\bf 272–276} 234–236

\bibitem{Kim03}
Kim M S, Echizen Y, Umeo K, Kobayashi S, Sera S, Salamakha P S, Sologub O L, Takabatake T, Chen X, Tayama T, Sakakibara T, Jung M H and Maple M B 2003 {\it Phys. Rev. B} {\bf 68} 054416

\bibitem{tsunetsugu1997}
Tsunetsugu H, Sigrist M and Ueda K 1997 {\it Rev. Mod. Phys.} {\bf 69} 809--864

\bibitem{And61}
Anderson P W 1961 {\it Phys. Rev.} {\bf 124} 41

\bibitem{Spa97}
Doradzi\'nski R and Spa\l ek J 1997 {\it Phys. Rev. B} {\bf 56} R14239;
Doradzi\'nski R and Spa\l ek J 1998 {\it Phys. Rev. B} {\bf 58} 3293;
for brief review see:  Spa\l ek J and Doradzi\'nski R 1999 {\it Acta Phys. Polon. A} {\bf 96} 677;
Spa\l ek J 2000 {\it Acta Phys. Polon. A} {\bf 97}, 71

\bibitem{Sle06}
\'Slebarski A 2006 {\it J. Alloys Compds.} {\bf 423} 15;
\'Slebarski A and Spa\l ek J 2007 {\it J. Magn. Magn. Mater.} {\bf 310} e85

\bibitem{sle2008}
\'Slebarski A, Maple M B, Baumbach R E and Sayles T A 2008 {\it Phys. Rev. B} {\bf 77} 245133

\bibitem{sle2009}
\'Slebarski A and Fija\l kowski M 2009 {\it Physica B} {\bf 404} 2969--2971

\bibitem{tou2004}
Tou H, Kim M S, Takabatake T and Sera M 2004 {\it Phys. Rev. B} {\bf 70} R100407

\bibitem{smith69}
Smith T F, Chu C W and Maple M B 1969 {\it Cryogenics} {\bf 9} 53

\bibitem{note}
Initial characterizations of the samples were performed at atmospheric pressure (1 bar) prior to the high-pressure experiments, for the same samples and the same leads used later in the pressure experiments, although the measurements were performed in a different cryostat than the one used for the high-pressure experiments. For this reason, electrical resistivity curves taken at 1 bar will not be compared with the curves measured at higher pressures.

\bibitem{Gamza09}
Gam\.za M, \'Slebarski A and Rosner H 2009 {\it Eur. Phys. J. B} {\bf 67} 483

\bibitem{sle2009b}
\'Slebarski A 2009 {\it J. Alloys Compds.} {\bf 480} 9–12

\bibitem{Hard80}
Hardebusch U, Gerhardt W and Schilling J S 1980 {\it Phys. Rev. Lett.} {\bf 44} 352;
Hardebusch U, Gerhardt W and Schilling J S 1985 {\it Z. Phys. B-Condensed Matter} {\bf 60} 463

\bibitem{Sch79}
Schilling J S 1979 {\it Adv. Phys.} {\bf 28} 657

\bibitem{note2}
In \cite{Sle06}, the total number of conduction electron states per atom $n_c \sim 1$ was calculated counting 1/2 state per $s$-electron, 1/6 state per $p$-electron, 1/10 state per $d$-electron, and 1/14 state per $f$-electron per atom. For CeRhSn, the authors considered the Ce atoms to be in the [Xe] 4f$^1$5d$^1$6s$^2$ state (1/14$f$+1/10$d$+2/2$s$ states/atom), the Rh atoms to be in the [Kr]4d$^8$5s$^1$ state (8/10$d$+1/2$s$ states/atom), and the Sn atoms in the [Kr]4d$^{10}$5s$^2$5p$^2$ state (2/2$s$+2/6$p$ states/atom). Based on XPS experiments and calculations, it was estimated that the Rh $d$-states are located 4 eV below the Fermi level, so it was assumed that the Rh $d$-states do not contribute to the conduction band. The value of $n_f \sim 0.93$ was also obtained from X-ray photoemission spectroscopy (XPS).

\bibitem{Maple74}
Maple M B and Wohlleben D K 1973 {\it Magnetism and Magnetic Materials-1973} {\bf 18}, AIP Conf. Proc. (C. D. Graham, Jr., J. J. Rhyne, eds., AIP, New York) p. 447

\bibitem{Don77}
Doniach S 1977 {\it Physica B} {\bf 91} 231;
Doniach S 1977 {\it Valence Instabilities and Related Narrow Band Phenomena} (in: R.D. Parks Eds., Plenum, New York)

\bibitem{Mederle02}
Mederle S, Borth R, Geibel C, Grosche F M, Sparn G, Trovarelli O and Steglich F 2002 {\it J. Phys.: Condens. Matter} {\bf 16} 10731

\bibitem{oomi2004}
Oomi G, Kagayama T, Aoki Y, Sato H, Onuki Y, Takahashi H and M\^{o}ri N 2002 {\it J. Phys.: Condens. Matter} {\bf 16} 3385–-3400

\bibitem{Griffiths69}
Griffiths R B 1969 {\it Phys. Rev, Lett.} {\bf 23} 17;
Castro Neto A H, Castilla G and Jones B A 1998 {\it Phys. Rev. Lett.} {\bf 81} 3531

\end{thebibliography}
\end{document}